\documentclass[prl,superscriptaddress,showpacs,showkeys,twocolumn]{revtex4-1}

\usepackage{amsfonts,amsmath,amssymb,graphicx,epstopdf,verbatim}

\usepackage[english]{babel}

\begin{document}

\title{Optical properties of inhomogeneous metallic hydrogen plasmas}

\author{N. Van den Broeck}
\author{F. Brosens}
\author{J. Tempere}
\affiliation{Theory of Quantum and Complex systems, Universiteit Antwerpen, Belgium}

\author{I.F. Silvera}
\affiliation{Lyman Laboratory of Physics, Harvard University, Cambridge, Massachusetts 02138}

\date{\today}

\begin{abstract}
We investigate the optical properties of hydrogen as it undergoes a transition from the insulating molecular to the metallic atomic phase, when heated by a pulsed laser at megabar pressures in a diamond anvil cell. Most current experiments attempt to observe this transition by detecting a change in the optical reflectance and/or transmittance. Theoretical models for this change are based on the dielectric function calculated for bulk, homogeneous slabs of material. Experimentally, one expects a hydrogen plasma density that varies on a length scale not substantially smaller than the wave length of the probing light. We show that taking this inhomogeneity into account can lead to significant corrections in the reflectance and transmittance. We present a technique to calculate the optical properties of systems with a smoothly varying density of charge carriers, determine the optical response for metallic hydrogen in the diamond anvil cell experiment and contrast this with the standard results. Analyzing recent  experimental results we obtain   
$\sigma^{\textnormal{Drude}}_{\textnormal{DC}}
=(2.1 \pm 1.3) \times 10^3$ ($\Omega$ cm)$^{-1}$ for the conductivity of metallic hydrogen at 170 GPa and 1250 K.
\end{abstract}  


\maketitle
\section{Introduction}

Metallic hydrogen (MH) has been a challenge for experiment and theory
since the initial proposal by Wigner and Huntington that
under sufficient pressure hydrogen would dissociate to 
an atomic metallic solid \cite{WignerJChemPhys1935}. 
In the past decades there have been a number of reports
of producing metallic hydrogen in the laboratory 
\cite{kawaiProfJapanAcad1975,VereshchaginJETPL1975,MaoScience1989,
MaoPRL1990, WeirPRL76,NellisPRB59, EremetsNatMat2011,DzyaburaPNAS2013,Zaghoo},
but only a shock experiment \cite{WeirPRL76,NellisPRB59}
and a recent static pressure experiment \cite{Zaghoo}
have provided convincing evidence of metallization.
Many of the experiments
use optical properties (when possible) to determine the state of the
hydrogen, as this is a non-intrusive technique. 
In experiments where transmittance and reflectance are measured, the interpretation of optical properties is based on treating the sample as a uniform metallic slab. Also current theoretical predictions \cite{CollinsPRB2001,HolstPRB2008} of the reflectance of MH are derived for a block of uniform density. 

In reality the samples are inhomogeneous conductors 
because the hot part of the sample is bounded by low 
temperature non-metallic regions of the same material.
Here we show that the slab model introduces inaccuracies in the analyses of optical properties for an inhomogeneous plasma density and we propose a theoretical description that remedies this.
Wave propagation through inhomogeneous plasmas has been widely studied, but here we no longer assume that the inhomogeneity is on a length scale longer than the wavelength, and focus on the opposite case. 
%
We introduce an algorithm that also allows the presence of other materials 
(e.g. a tungsten absorber or a cladding layer) to be taken into account 
and that allows the effect of edge smoothing or diffusion of materials into 
each other to be investigated. The model is applicable
to any situation where inhomogeneous plasma densities can occur, 
and thus is also relevant in, for example, studies of fusion plasmas, 
plasma materials processing reactors, and the ionosphere. 
Combining our method with the recent
measurements of Zaghoo, Salamat and Silvera 
in a pulsed-laser heated diamond anvil cell
(Ref. \cite{Zaghoo}, hereafter referred to as ZSS),
we estimate the conductivity of MH, a quantity that is of 
importance in planetary science as high-pressure 
high-temperature hydrogen is a major component 
of gas giants.

\section{Theoretical description}

We consider a sample with charge carriers (here electrons) with a density
profile $n(\mathbf{r})$. When \emph{monochromatic}
light of frequency $\omega$ illuminates the sample, the total electric field
$\mathbf{E}(\mathbf{r},\omega)$ causes the electrons to be displaced out of
equilibrium over a distance $\mathbf{X}(\mathbf{r},\omega)$ that must be
calculated in general from a microscopic response theory. Here, since the
subject of our paper is not the microscopic theory but rather the macroscopic
inhomogeneity in $n(\mathbf{r})$, we will use a generalisation of the Drude response given in SI units by
\begin{equation}
 \mathbf{X}(\mathbf{r},\omega)=\frac{e}{m_{e}}
 \frac{\mathbf{E}(\mathbf{r},\omega)}{\omega(\omega+i/\tau)},
\end{equation}
where $e$ and $m_{e}\,$are the electron charge and mass, respectively, and
$\tau$ is the Drude relaxation time. The displacements $\mathbf{X}(\mathbf{r}%
,\omega)$ lead to induced charge and current densities given by
\begin{align}
 \rho^{ind}(\mathbf{r},\omega)=-e\nabla\cdot\left[  
   n(\mathbf{r})\mathbf{X}(\mathbf{r},\omega)\right], \\
 \mathbf{J}^{ind}(\mathbf{r},\omega)=i\omega en(\mathbf{r})
   \mathbf{X}(\mathbf{r},\omega).\label{jind0}%
\end{align}
It is useful to rescale the density profile $n(\mathbf{r})$ by the bulk free
electron density $n_{0}$, introducing $f(\mathbf{r})=n(\mathbf{r})/n_{0}$. The
induced charge density can then be rewritten using the definition of relative
permittivity of a Drude metal, 
$\epsilon(\omega)=1-\omega_{pl}^{2}/[\omega(\omega+i/\tau)]$ where
$\omega_{pl}=\sqrt{n_{0}e^{2}/(m_{e}\varepsilon_{vac})}$ 
is the plasma frequency, $n_{0}$ the bulk free
electron density, and $\varepsilon_{vac}$ the vacuum permittivity:
\begin{equation}
  \rho^{ind}(\mathbf{r},\omega)=\sum_{j}\varepsilon_{vac}\left[ 1-\epsilon_{j}(\omega)\right]  \nabla\cdot\left[  f_{j}(\mathbf{r}) \mathbf{E}(\mathbf{r},\omega)\right].\label{rhoind}
\end{equation}
Here we also introduced an index $j$ in order to take into account the
presence of different materials. The profile function $f_{j}(\mathbf{r})$
defines the location of material $j$ and can describe smoothing of the edges
of the material. Similarly, the induced current density (\ref{jind0}) can be
rewritten as
\begin{equation}
 \mathbf{J}^{ind}(\mathbf{r},\omega)=\sum_{j}i\omega\varepsilon_{vac}\left[
 1-\epsilon_{j}(\omega)\right]  f_{j}(\mathbf{r})
 \mathbf{E}(\mathbf{r},\omega).\label{jind}%
\end{equation}
These induced charges and currents in turn induce electric and magnetic
fields. Note that in the above expressions, $\mathbf{E}(\mathbf{r},\omega)$ is
the total field, i.e. the sum of (i) the field induced by 
$\rho^{ind},\mathbf{J}^{ind}$, and (ii) the externally applied field (the light beam). 
Relations (\ref{rhoind}) and (\ref{jind}) can be used to eliminate the induced 
charge and current density from the Maxwell equations. Writing these in terms 
of the vector potential $\mathbf{A(r},\omega)$ and the scalar potential
$\phi\mathbf{(r},\omega)$ instead of electric and magnetic field, we obtain
the following set of equations:
\begin{equation}
 \Delta\phi^{ind}=\sum_{j}\left(  1-\epsilon_{j}\right)  \left[
 f_{j}\Delta\phi-\mathbf{\nabla}f_{j}\cdot\left(  -\mathbf{\nabla}\phi
 +i\omega\mathbf{A}\right)  \right]  ,  
 \label{eq:diffeqnPhi} 
\end{equation}
and
\begin{multline}
 \label{eq:diffeqnJ}
 \left(  \Delta+\frac{\omega^{2}}{c^{2}}\right)  \mathbf{A}^{ind} =
 i\frac {\omega}{c^{2}}\sum_{j}\left(  1-\epsilon_{j}\right)  f_{j}\left(
 \mathbf{\nabla}\phi-i\omega\mathbf{A}\right) \\
-i\frac{\omega}{c^{2}}\mathbf{\nabla}\phi .
\end{multline}
%
%
with $c$ the velocity of light in vacuum. Here $\phi^{ind}$ and $\mathbf{A}^{ind}$ represent the scalar and vector
potentials induced by $\rho^{ind}$ and $J^{ind}$, whereas $\phi$ and
$\mathbf{A}$ represent the total scalar and vector potentials, including also
the externally applied potentials. Note that we have left out the 
$\mathbf{r},\omega$ dependence for notational simplicity.
The Coulomb gauge $\nabla\cdot \mathbf{A}=0$ has been used.

In a diamond anvil cell (DAC), equations (\ref{eq:diffeqnPhi})-(\ref{eq:diffeqnJ}) are simplified by taking the uniaxial geometry of the system into account.
Here, we take the $x$-axis to be aligned along the optical axis of the diamond anvils, and assume homogeneity in the $yz$ plane.
The incoming light is considered to be a plane wave travelling 
along the $x$-axis, corresponding to $\phi^{ext} = 0$ and 
$\mathbf{A}^{ext}\left(x,\omega\right) = A_0e^{i\left(kx-\omega t\right)}\mathbf{e}_y $.
When using this in the differential equations (\ref{eq:diffeqnPhi}) and
(\ref{eq:diffeqnJ}), it is readily seen that
$\phi^{ind}=A^{ind}_z=A^{ind}_x=0$. 
The only non-zero component $A^{ind}_y$ can be found from the remaining 
differential equation:
\begin{multline}
 \label{eq:diffeqAy}
 \left(\Delta+\sum_j k^2\left[1-\left(1-\epsilon_j\right)f_j\right]\right) A^{ind}_y \\
 = \sum_jk^2\left(1-\epsilon_j\right)f_jA_0e^{i\left(kx-\omega t\right)}.
\end{multline}
This equation is solved numerically on a grid of $N$ points with spatial divisions $\delta x$ (a non-uniform grid is also possible). Using the central finite difference method, the differential equation can be rewritten as the recursion relation
\begin{multline}
 \label{eq:recurRel}
 u_{\alpha+1} = \textstyle{\sum_j} \left(1-\epsilon_j\right)
  f_j\left(x_\alpha\right)e^{ikx_\alpha}k^2\delta x^2
 \\
 +\left[2-\left(1-\textstyle{\sum_j} \left(1-\epsilon_j\right)f_j
  \left(x_\alpha\right)\right)k^2\delta x^2\right]u_\alpha-u_{\alpha-1}
\end{multline}
where $u_\alpha$ indicates the solution for $A_y^{ind}/A_0$ on grid point $\alpha$. When the values of the two initial grid points $u_1$ and $u_2$ are known, this recursion relation can be used to calculate the remaining points. If we take the incoming wave to start from $x_1$ and propagate towards $x_N$, then the values of the induced solution at the first two points will be the reflected wave.
This statement assumes that the first grid points are located in the surrounding medium outside the system under consideration.
In this case $u_1=C_R$ and $u_2=C_R e^{-ik\delta x}$, where $C_R$ is an undefined complex constant connected to the reflectance through $R=\left|C_R\right|^2$. Similarly, we define $C_T$ from the last grid point, again located in the surrounding medium, $u_N\left(C_R\right)=C_Te^{ikx_N}$, and relate it to the transmittance coefficient through $T=\left|1+C_T\right|^2$, where the term $1$ comes from the external wave. 
Furthermore, we must demand that the solution at $u_N$ and $u_{N-1}$ is in fact an outgoing wave of the form $e^{ikx}$. Hence we must impose
$ u_{N-1}\left(C_R\right)e^{-ik\delta x} = u_N\left(C_R\right)$.
The goal of the algorithm will thus be to minimize
\begin{equation}
 \label{eq:RVWalgo}
 \left|u_{N-1}\left(C_R\right)e^{-ik\delta x} - u_N\left(C_R\right)\right|
\end{equation}
\noindent with respect to $C_R$. The solution yields the correct $C_R$ and $C_T$ from which the reflectance $R$ and transmittance $T$ can be calculated. In turn, these two variables easily provide the absorption $A=1-R-T$ and thus completely define the optical response of the system.

\section{Effect of inhomogeneity}

When considering a semi-infinite block of material with
sharp interface to the vacuum and permittivity $\epsilon(\omega)$, 
the reflectance is given by the textbook result 
$R_0(\omega)=| (\sqrt{\epsilon}-1)/(\sqrt{\epsilon}+1) |^2$ .
This formula has been used in Refs. \cite{CollinsPRB2001,HolstPRB2008} 
in conjunction with microscopic derivations of the permittivity 
$\epsilon(\omega)$ of metallic hydrogen to predict the 
change in reflectance upon metallization.
In order to demonstrate that $R_0(\omega)$ can give an incorrect result, and the formalism presented above provides a different result appropriate for an inhomogeneous density profile, we compute the reflectance for a density profile modeled by
\begin{equation}
 \label{eq:profileSlab}
  f\left(x\right) = \frac12 \frac{\exp\left(\frac{d}{2\kappa}\right)}
  {\cosh\left(\frac{d}{2\kappa}\right)+\cosh\left[ (2x-\frac{d}{2}) / \kappa \right]}.
\end{equation}
\begin{figure}
	\centering
	\includegraphics[scale=0.6]{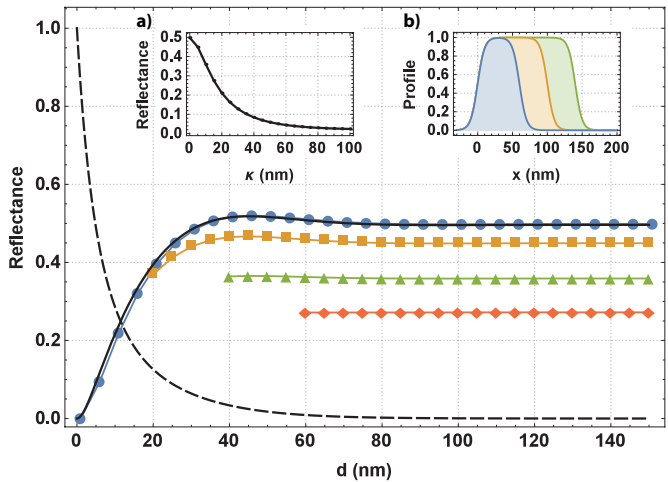}
	\caption{(Color online) The reflectance of the density profile presented in eq. (\ref{eq:profileSlab}) for MH in function of the thickness $d$ for $\kappa = $ 0, 5, 10 and 15 nm (disks, squars, triangles and diamonds resp.). The solid line coinciding with the disks represents the analytical result for a hard wall. The black dashed line shows the analytical transmittance. Transmittance for different $\kappa$ is not shown since they all coincide with the dashed line. Data points with combinations of $d$ and $\kappa$ resulting in a density profile that does not reach 1 have been omitted. Inset (a) shows the reflectance as a function of $\kappa$ for $d=500$ nm. In this case, transmittance is zero for all $\kappa$. Inset (b) shows the density profile for $d =$ 60, 100 and 140 nm with $\kappa = 5$ nm. The parameters used in all these figures are: permittivity $\epsilon=-1.47+13.6 i$ and $\lambda=500$ nm.}
	\label{fig:square}
\end{figure}
This represents a layer of thickness $d$, with edges smoothed over a distance $\kappa$, as illustrated in inset (b) of Fig. \ref{fig:square}.
In the limit $\kappa \rightarrow 0$ we retrieve a sharp-edged layer. In 
Fig. \ref{fig:square} the reflectance for such a smooth-edged layer is shown
as a function of increasing thickness $d$, 
keeping $\kappa$ fixed (at 0, 5, 10 and 15 nm for disks, squares, triangles and diamonds respectively). As the layer thickness grows, the reflectance goes 
to a constant value, but only for $\kappa \rightarrow 0$ does this correspond to the 
result $R_0(\omega)$ for a sharp interface, 
shown as a full curve. As the interface is smoothed,
the reflectance decreases in favor of the absorptance. This decrease is also
shown in inset (a). This means that even for thick films, the reflectance is
influenced strongly by the smoothness of the interface.
The enhancement of absorption can be related to a smooth changing 
permittivity, which is comparable to techniques used in graded 
refractive index antireflection coatings \cite{MahdjoubTSF2005}.

\begin{figure}
	\centering
	\includegraphics[scale=0.7]{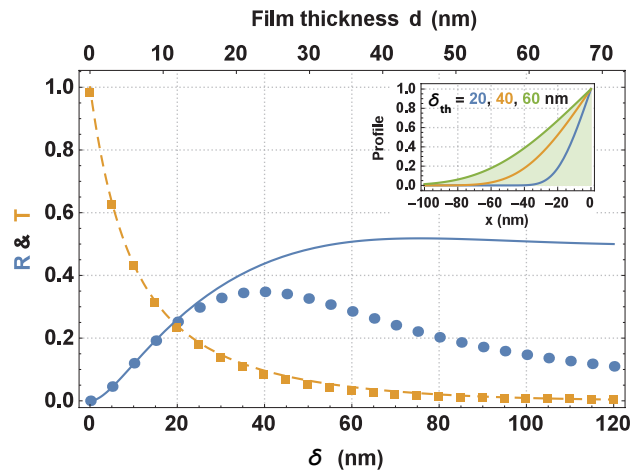}
	\caption{(Color online) Reflectance and transmittance are shown as a function
	of layer thickness for a metal with $\epsilon=-1.47+13.6 i$. 
	The solid (reflectance) and dashed (transmittance) curves show the result for a homogeneous slab of thickness $d$, and
	the disks show the result for the inhomogeneous density profile $f(x)$ shown
	in the inset and representing a smooth increase from zero to maximum density
	over a distance $\delta$, after which the density is zero (i.e. for $x>0$).
	Light with wavelength $\lambda=500$ nm is incident from $x=-\infty$.}
	\label{fig:triangle}
\end{figure}
In Fig. \ref{fig:triangle}, results are shown for a different kind of
density profile. In pulsed-laser heated DACs one expects the 
temperature to drop off from a maximum at the heater to zero 
smoothly \cite{RekhiRSI2003}, over a distance $\delta$. A 
similar profile can then be expected for the density of 
dissociated molecules. This type of profile, illustrated in the inset 
of Fig. \ref{fig:triangle} and parametrized by 
$f(x)=\exp[1-\exp(-x/\delta)]$ for $x\le0$, is used
to plot the reflectance and transmittance as 
a function of $\delta$ in Fig. \ref{fig:triangle}.
The disks and
squares show the results with the current formalism, and the solid and dashed
curves show the analytical result for a sharp-edged layer with the same total amount of charge carriers.
For thin films, the precise shape of the density profile is unimportant since the layer becomes effectively a two-dimensional sheet of charge carriers.
However, as the thickness is increased, we again find that the inhomogeneity of the density profile leads to a reduction of reflectance in favor of absorption.
Hence, it is necessary to take the inhomogeneity into account when inferring the film thickness from an optical reflectance measurement, once the film becomes thicker than a few
tens of nanometer.
\section{Metallic hydrogen in a DAC}
In ZSS the phase transition to liquid metallic hydrogen was 
observed in a DAC by heating the sample along an isochore. 
A thin film of tungsten is used to heat the hydrogen by 
absorbing a few hundreds nanosecond long laser pulse \cite{RekhiRSI2003}. 
The W film has been constructed to absorb 50\% of the incoming light and transmit another 25\%, which corresponds to a film thickness of 8 nm for a laser of $\lambda=500$ nm.
We take the frequency dependent tungsten permittivity from tables published in \cite{CRC}.
The tungsten film is separated from the hydrogen by a 5 nm cladding layer of (optically inactive) alumina.

As the pulsed laser power is increased, the temperature of the 
tungsten film grows, and when the 
temperature $T$ reaches the dissociation temperature 
$T_d$, a layer of metallic hydrogen is formed, reducing transmittance 
and increasing reflectance from the sample.
The density of the metallic hydrogen is expected to be highest where it is in contact with the tungsten film. 
Combining the Arrhenius equation \cite{ArrheniusZPC1889} for the 
dissociation rate of molecules, and the temperature profile for 
pulsed laser heating derived in \cite{RekhiRSI2003}, we propose
the following density profile for the metallic hydrogen:
\begin{equation}
 \label{eq:profile2H}
 f_{H}\left( x \right) = \exp \left[-(T_{d}/T) e^{x / \delta_{th}} \right],
\end{equation}
The maximum number density of metallic hydrogen is 
taken to be twice the molecular hydrogen density 
obtained from the equation of state \cite{hydrogenRMP}. 
The maximum MH density occurs at $x=0$, close to the 
tungsten layer, and drops off for $x>0$. 
In the region $x<0$, the density of 
metallic hydrogen is zero, and for the remainder of the 
calculations we set $T=T_d$. The resulting density profile 
(\ref{eq:profile2H}) is similar to those illustrated in 
the inset of Fig. \ref{fig:triangle}.

The theoretical description introduced above enables us to investigate
the difference in reflectance (transmittance) between this inhomogeneous density profile and the reflectance (transmittance) obtained for a uniform slab of MH.
This is of importance in order to correctly derive the
properties of the newly detected phase from the observed
changes in optical response.
From Ref. \cite{CollinsPRB2001} we have $\tau = 4.8\times10^{-17}$ s 
and $\omega_{pl}=22$ eV at 30000 K and 152 GPa, resulting in $n_0=3.5\times10^{29}$ m$^{-3}$ and a permittivity $\epsilon_{MH}=-1.47+13.6 i$ at $\lambda = 500$ nm as modelled by $\epsilon(\omega)=1-\omega_{pl}^{2}/[\omega(\omega+i/\tau)]$.
\begin{figure}
	\centering
	\includegraphics[scale=0.8]{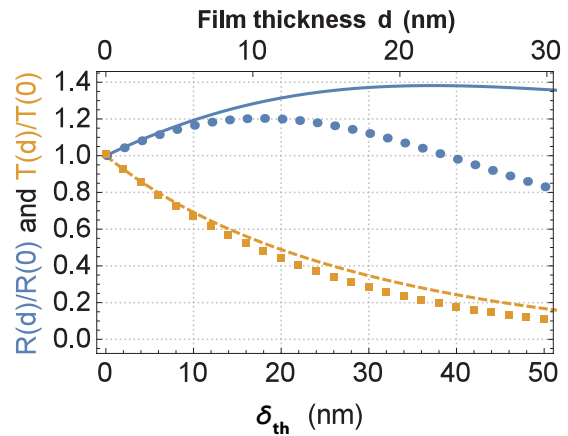}
    \caption{(Color online) The change in reflectance and transmittance 
    at $\lambda=500$ nm, is shown as 
    a function of the thickness $d$ of a layer of metallic hydrogen
    on top of a 15 nm thick layer of tungsten. The disks and squares show 
    the reflectance and transmittance, respectively, for an inhomogeneous 
    density profile as given by eq. (\ref{eq:profile2H}). The solid and dashed line show reflectance and transmittance, respectively,
    for a layer of hydrogen with uniform density. The thickness of the 
    uniform layer is chosen such that the total amount of electrons 
    is the same as that for the non-uniform profile, and it is 
    shown on the upper axis.}
    \label{fighydro}
\end{figure}

Figure \ref{fighydro} contrasts the results for a uniform layer of
hydrogen with that of a density profile as given by eq. (\ref{eq:profile2H}).
The change in reflectance (transmittance) is calculated as the ratio of the 
reflectance (transmittance) with a hydrogen layer of thickness $\delta_{th}$
and without the hydrogen layer ($\delta_{th} \rightarrow 0$). Results for a
uniform hydrogen density are shown as a solid curve (dashed curve),
and results for the non-uniform case are shown as disks (squares).
The non-uniform layer of the thickness $\delta_{th}$ is compared
to a uniform film with thickness $d$ chosen such that the density
integrated over $x$ is the same. Discrepancies start appearing
already at ca. 10 nm thickness, and are more pronounced for the
change in reflection. This is due to the fact that the inhomogeneous
density profile leads to a stronger absorption, which may even
result in a relative decrease of the reflectance for thicker
layers. In practice, this increased absorption by the metallic
hydrogen layer will lead to an increase in temperature above
the dissociation temperature. 

We can use these results to analyse the optical measurement 
of ZSS. We use the data at 170 GPa, namely the 
reflectance change at 980 nm and the transmittance change at 
980 nm and 633 nm, and focus on data in the temperature 
range 1260-1290 K just above the heating plateau that
indicates a first order transition to the 
plasma phase \cite{Zaghoo}. We set 
$\omega_{pl} = 26$ eV, corresponding to
a charge density twice the hydrogen molecular 
density at 170 GPa, and use a fit to the
observed reflectance and transmittances to 
obtain a Drude relaxation rate 
$\tau = (1.1 \pm 0.4)\times 10^{-17}$ sec.
The corresponding Drude conductivity is
$\sigma^{\textnormal{Drude}}_{\textnormal{DC}}=(2100\pm1300)$
($\Omega$ cm)$^{-1}$. This value is in agreement with 
the value of 2000 ($\Omega$ cm)$^{-1}$ found for fluid hydrogen 
in shock wave experiments \cite{NellisPRB59} 
at 140 GPa and 2600 K, even though the
shock experiment finds a continuous transition
from a semiconducting to a metallic fluid, whereas
ZSS observe a first-order phase transition.
This may indicate that in both cases the metallic fluid
is very disordered, and has a conductivity
close to the Mott-Ioffe-Regel 
minimum metal conductivity\cite{Mott}. The relaxation
rate corresponding to the minimum metallic conductivity
in two dimensions is $\tau_{min}=0.68\times10^{-17}$ sec. 
The film thickness compatible
with our fit is 1-3 nm, which also hints at the
2D nature of the MH film formed in the experiment.
Finally, note that changing the plasma frequency 
to 21 eV (the value reported in \cite{CollinsPRB2001})
would yield $\tau = (1.7 \pm 0.9)\times 10^{-17}$ sec.
%
\section{Conclusions}
In this article we propose a method to obtain the reflectance,
transmittance and absorptance for an inhomogeneous charge density,
given the microscopic response (i.e. the dielectric function). 
The results are contrasted to results for uniform 
slabs of materials obtained with the standard Maxwell boundary 
conditions for sharp interfaces. The theoretical description
is then applied to the case of liquid metallic 
hydrogen \cite{Zaghoo} in a DAC, 
where it is found that taking this into account reduces the 
reflectance of the metallic hydrogen with respect to that 
obtained in previous works for thick samples 
\cite{CollinsPRB2001,HolstPRB2008},
and increases the absorptance. Finally, we extract 
the conductivity of metallic hydrogen from the measurement
data of ZSS. 
The theory presented here will be important for 
forthcoming experiments that attempt to refine 
the optical measurements used to probe the newly detected 
metallic plasma phase of hydrogen.

\begin{acknowledgments}
Acknowledgments -- The authors would like to thank M. Zaghoo, and A. Salamat for discussions and suggestions. Part of this research was funded by a Ph.D. grant of the Agency for Innovation by Science and Technology (IWT). This research was supported by the Flemish Research Foundation (FWO-Vl), project nrs. G.0115.12N, G.0119.12N, G.0122.12N, G.0429.15N and by the Scientific Research Network of the Research Foundation-Flanders, WO.033.09N. This research was also supported by the NSF, grant DMR-1308641, and the DOE Stockpile Stewardship Academic Alliance
Program, grant DE-FG52-10NA29656.
\end{acknowledgments}

\end{document}